# Heart Signal (ECG) as Vibration Signature for Human Authentication Using TinyML

Yogeswar Reddy Thota*, Student Member, IEEE*, Hanan Alsmadi, *Student Member, IEEE*
and Tooraj Nikoubin, Senior Member, IEEE

***Abstract:*** **Ensuring data security in the era of Artificial Intelligence (AI) and deepfake technology requires robust biometric systems. Conventional methods such as facial recognition, iris scans, fingerprints, and voice recognition are increasingly vulnerable to spoofing and provide only one-time authentication at login. This study introduces an Electrocardiogram (ECG) based authentication framework, compared with Photoplethysmography (PPG), leveraging Tiny Machine Learning (TinyML) for deployment on resource-constrained microcontroller units (MCUs). The proposed method enables continuous authentication, offering ongoing verification throughout device usage, which conventional biometrics cannot achieve. It also serves as a password-free solution, reducing risks of credential loss, theft, or reuse. However, ECG signals vary with physical activity, health conditions, age, and emotional states, requiring robust signal processing and learning algorithms. To address this, efficient TinyML models were designed to mitigate recognition errors from noisy real-time data. Experiments used single-lead ECG signals (Lead I) from UofTDB, MIT-BIH Arrhythmia, MIT-BIH Noise Stress Test, and PTB databases. Spectrograms and spectral features were extracted to train Deep Neural Network (DNN), 1D CNN, and 2D CNN models within the Edge Impulse platform, achieving 98% accuracy. The trained model was converted into TinyML format and deployed on an Espressif ESP-EYE (ESP32, 240 MHz). Support Vector Machine (SVM) models were also applied for benchmarking. By analyzing the number of heartbeats required for reliable recognition, results show that averaging multiple signals improves tolerance to noise and enhances accuracy. Overall, ECG-based authentication demonstrates strong potential as a secure, real-time, and continuously verifiable method for personal identification on wearable and edge devices.**

Yogeswar Reddy Thota, is with the Electrical and Computer Engineering Department, University of Texas at Dallas, EC 33, 800 West Campbell Road, Richardson, Texas 75080, USA. yogeswarreddy.thota@UTDallas.edu

Hanan Alsmadi, is with the Electrical and Computer Engineering Department, University of Texas at Dallas, EC 33, 800 West Campbell Road, Richardson, Texas 75080, USA. hanan.alsmadi@UTDallas.edu

Tooraj Nikoubin, is with the Electrical and Computer Engineering Department, University of Texas at Dallas, EC 33, 800 West Campbell Road, Richardson, Texas 75080, USA Tooraj.Nikoubin@UTDallas.edu

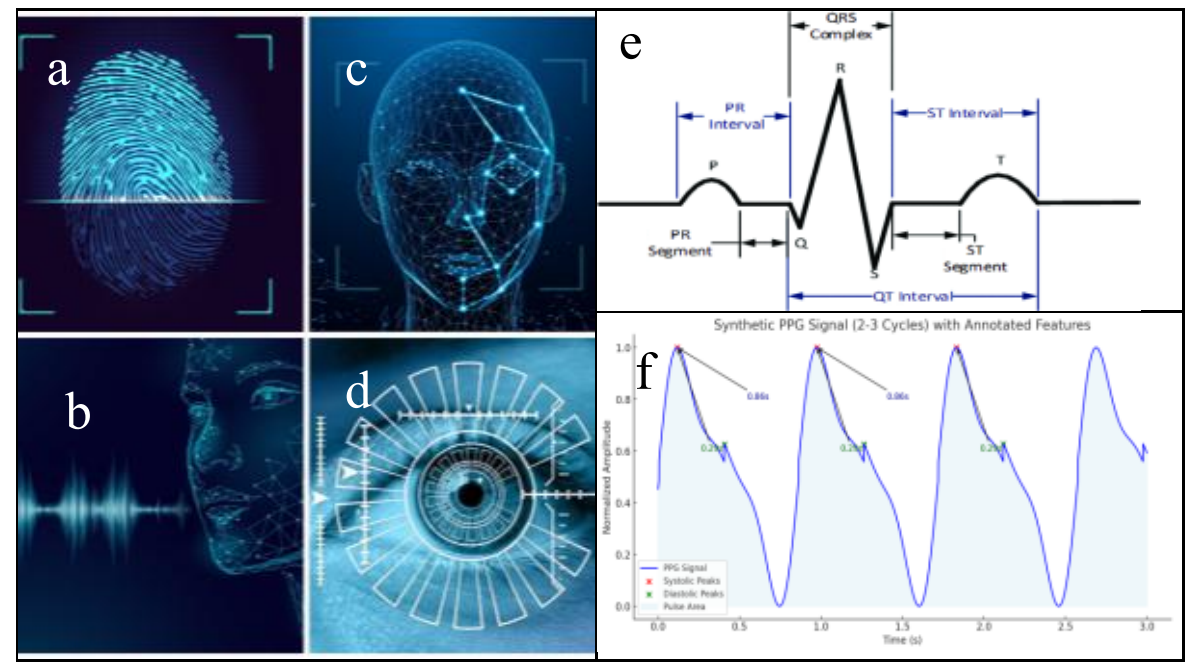

*Fig 1: Traditional biometrics and physiological signals (ECG and PPG).*

## I. Introduction

OVER the past several years, numerous studies have explored the potential of using Electrocardiogram (ECG) and Photoplethysmography (PPG) signals as a biometric authentication system. As time has passed and technology has advanced, the traditional biometric methods such as fingerprint, facial recognition, voice, and iris scanning (Fig. 1) have become susceptible to forgery and misuse [1]–[3]. There is thus a need for an alternative biometric that offers better security and protection against counterfeiting. Because of its universality, performance, uniqueness, liveliness detection, and robustness to attacks, ECG, PPG signals has become a strong alternative for biometric systems [4]. Continuous biometric authentication has gained attention as a next-generation technique due to the unique characteristics of these physiological signals, making it a promising trait for continuous authentication [5]. In [6, 7], results show that extracted features are independent of sensor location and unaffected by individual anxiety levels, affirming the robustness of ECG signals for biometric authentication purposes. However, due to the relatively recent use of ECG as a biometric trait and the lack of extensive databases, ECG signal processing still faces several challenges [4]. Analyzing these signals requires multiple steps, including feature extraction, filtering, segmentation, and matching with signals in the database. These steps are detailed later in the paper. Different components of an ECG signal are shown in Fig. 1(e), and heartbeat signal formation is illustrated in Fig. 2.

The Tiny Machine Learning (TinyML) paradigm revolutionizes the integration of Machine Learning (ML) capabilities into small, Microcontroller Unit (MCU)-powered devices. Cloud-based processing requires constant internet access, leading to higher power consumption and vulnerabilities from data transmission. By eliminating this dependency, TinyML enables innovative applications with improved power efficiency and privacy [9]. TinyML allows intelligent processing within embedded, resource-constrained devices. Its applications span from healthcare to transportation, and in this paper, we focus on security applications, leveraging TinyML's ability to process ECG, PPG signals for authentication [10, 32]. Alongside the rise of the Artificial Intelligence of Things (AIoT), TinyML has become an important paradigm [9]. TinyML-based edge computing plays a pivotal role in enabling real-time decisions for authentication using ECG or PPG biometric data captured at the source. This approach bypasses vulnerabilities introduced by cloud dependence, making it less susceptible to tampering or cyberattacks. By processing data locally in offline mode and providing immediate feedback, it significantly reduces latency and power usage. Instead of sending data to and from centralized systems, TinyML enables immediate computation through on-device ML models.

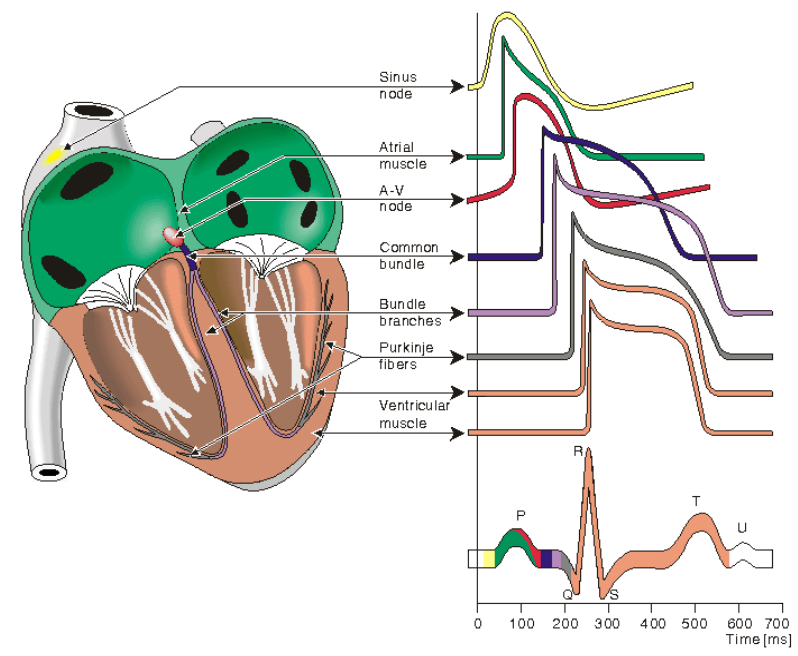

Fig. 2. Heart pulse formation through cardiac activity [8].

### *A. Background and Philosophy of the work*

The promising characteristics of ECG as biometrics opened doors for many researchers to study and introduce new reliable methodologies to achieve this. In [11], authors used waveforms divided by the fiducials (WFs) features to test Neural Network (NN) and Support Vector Machine (SVM) performance in MATLAB®, resulting in 96.6% and 97.6% accuracy for 85 subjects. These features require more computational power to train on a large dataset. In [12], authors used Shannon energy, skewness, occupied bandwidth and other statistical features to train the SVM with a cubic kernel model for 14 subjects' ECG data filtered by empirical mode decomposition (EMD), which increased the complexity and achieved 98.7% accuracy in MATLAB®. In [13], TinyML-based continuous ECG monitoring was introduced to detect arrhythmia with approximately 97% accuracy using the Convolutional Neural Network (CNN) model. This work shows that TinyML has great potential for processing physiological signals like ECG. In [14], the authors implemented a CNN model trained using spectrograms and achieved a maximum of 99% accuracy on 40 subjects. However, training was performed using two NVIDIA GeForce GTX 1080 Ti GPUs, requiring a high-end hardware setup. In contrast, the proposed model in this paper achieves similar validation using a Deep Neural Network (DNN) model deployed through Edge Impulse on a TinyML hardware module. Some studies have explored ECG biometrics on mobile platforms. For example, Arteaga-Falconi et al. [31] proposed an ECG authentication system for mobile devices, using fiducial feature segmentation and classifier models. Similarly, Sidek et al. [22] used data mining methods for ECG biometric identification in mobile environments. However, these mobile-based approaches still rely on powerful processors or partial offloading to cloud services, and do not support fully embedded, resource-constrained edge devices.

In contrast, as per our knowledge this work is the first to demonstrate

- End-to-end ECG-based biometric authentication using TinyML on an low-power MCU (ESP32, 240MHz), where model training and inference are optimized for real-time, on-device operation without any cloud dependency.
- This lightweight implementation can seamlessly blend into wearable devices that operate on limited energy and processing power, enabling continuous authentication without draining resources.
- Unlike mobile-based systems that often rely on high-performance processors or intermittent connectivity, our approach is designed to function efficiently on always-on, battery-powered devices making it suitable for practical, real-world deployment in smartwatches, rings, fitness trackers, and other AIoT applications.

In our previous work [29], we evaluated the SVM model with statistical and time domain features and achieved an accuracy of 87%, while the DNN model with spectrogram features reached 97%. We also evaluated different DNN architectures to determine which are best suited for TinyML devices, In this paper we introduced,

- Optimized biometric authentication through ECG signals by evaluating features across different categories and applying ML models (SVM, DNN, 1D CNN, 2D CNN) tailored for resource-constrained edge devices.
- Comparison of ECG with PPG signal processing and how ECG signal have more parameters than PPG resulting in better performance of the authentication, especially when it goes to large scale authentication applications.
- Model performance evaluation under varying sampling frequencies, across sessions, and under different conditions with arrhythmia and added noise.
- Training efficiency analysis by determining the minimum number of ECG segments required for effective model performance, achieving reliable authentication with minimal input data for real time applications.

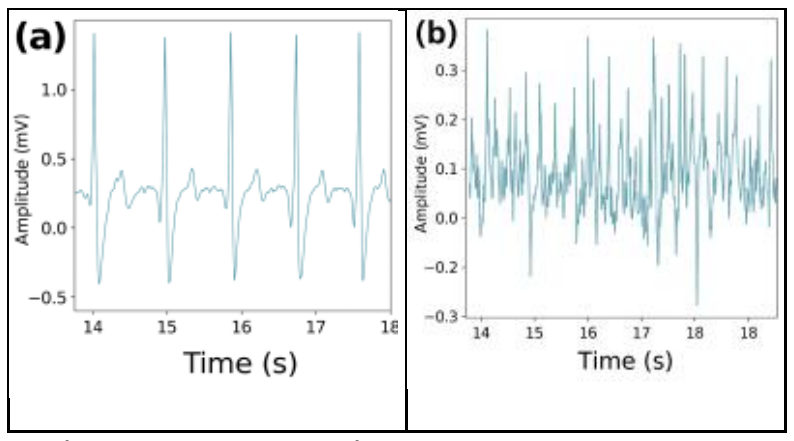

Fig. 3. ECG at a) Resting and b) Exercising states

- Cross-database validation to confirm the robustness and generalizability of our models across multiple datasets with diverse subject profiles.
- Identification space compression: By segmenting ECG around R-peaks and extracting discriminative features (e.g., spectrograms), our approach maps each user's heartbeat into a low-dimensional yet highly distinctive representation. This enables faster classification and reduces memory footprint on embedded systems.
- On-device real-time authentication: Our fully deployed TinyML pipeline on an ESP32 MCU supports continuous identity verification without cloud dependency, proving the feasibility of privacy-preserving, always-on biometric systems for wearable's and IoT healthcare applications.
- Comparison of ECG with PPG signals processing for biometrics and their performance evaluation for future biometric sensing design platforms.

The paper is further organized as follows: Section II presents the methodology; Section III details the optimization techniques and results; Sections V and VI discuss observations and conclusions, respectively.

## II. Methodology

In this research, Lead I ECG data was collected from some of the popular databases. After pre-processing these ECG signals, different categories of features were extracted from the actual segmented ECG signal and the spectrogram version of the segmented signal. An SVM model was trained in Python on a personal computer with a 64-bit operating system, Intel(R) Core(TM) i5-8250U CPU @ 1.60GHz 1.80GHz processor with 8.00GB RAM. The same preprocessed data was also used to train DNN, 1D CNN, and 2D CNN models in Edge Impulse. Later, the proposed model was compiled using TensorFlow and deployed onto the Espressif ESP-EYE (ESP32 240MHz) microcontroller, which supports TinyML operations remotely. The on-device performance metrics measured from this deployment are presented in Table V. The detailed methodology followed for both SVM and NN model analysis is explained below, and the proposed flow that achieved the best performance in the study is illustrated in Fig. 9, which is discussed further in this section

### *A. Databases*

In this analysis, the PTB Diagnostic [15], MIT-BIH Arrhythmia [16], MIT-BIH Noise Stress Test [17] from the PhysioNet database [18], and the University of Toronto ECG database (UofTDB) [19] were used to evaluate the model's performance under various conditions. These databases were chosen to represent multiple signal acquisition setups, activity types, and health conditions. The distribution of subjects across different physical states and sessions (i, ii, iii, iv, v, vi) from the UofTDB is detailed in Table I. This table provides insight into subject coverage across resting, exercising, and transitional conditions, which is essential for testing the robustness of ECG-based biometric models. ECG variance between resting and exercising states is illustrated in Fig. 3, showing how heart rhythm characteristics change under different physiological conditions. Such variation is essential in testing whether the models maintain classification performance across stress-induced signal changes.

TABLE I
UofTDB Subjects Distribution

| # Subjects at each recorded sessions | | | | | | #Subjects at each condition | | | | |
|---|---|---|---|---|---|---|---|---|---|---|
| i | ii | iii | iv | v | vi | a | b | c | d | e |
| 1020 | 72 | 65 | 54 | 47 | 43 | 81 | 1020 | 63 | 63 | 71 |
| 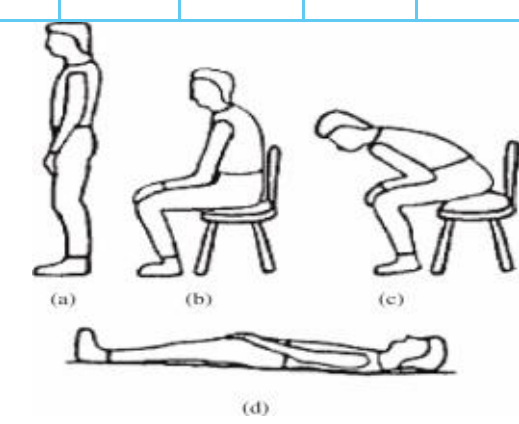 | | | | | | Age Range: 18-52, Gender: 61% Female, 39% male, Total number of subjects =1020.<br>Conditions:<br>(a) Stand (b) Sit (c) Tripod<br>(d) Supine (e) Exercise | | | | |

### *B. Pre-Processing*

Raw ECG signals are inherently susceptible to noise and artifacts, which can significantly affect authentication accuracy. Various factors, such as muscle activity or electromyography (EMG noise), baseline wander, electrode contact inconsistencies, and power line interference, similarly for PPG motion artifacts, baseline drift, sensor inconsistencies, and ambient light interference, introduce distortions that obscure key biometric features. To ensure clean and reliable signals, several preprocessing techniques were applied A comparative table summarizing the preprocessing steps for both ECG and PPG signals is presented in Table 2.

#### *1) Filtering the raw ECG signals*

In our analysis, for the ECG signal database with a sampling frequency of 200 Hz, a 4th-order Butterworth bandpass filter with cutoff frequencies of 0.5 Hz and 40 Hz [20] was applied to remove low-frequency artifacts like baseline wander (very low-frequency noise), which can be caused by factors like breathing, body movement, or electrode displacement, and high-frequency noise preserving the essential ECG features. This filter was chosen because it allows the key components of the ECG signal (P-wave, QRS complex, and T-wave) to pass through while removing baseline wander and high-frequency noise. For databases with higher sampling frequencies (e.g., 360 Hz, 1000 Hz), a 4th-order Butterworth low-pass filter with a cutoff frequency of 40 Hz [21] was used to attenuate high-frequency noise while preserving the key features of the ECG signal.

#### *2) Segmentation of Filtered signals*

ECG segmentation is the most popular signal preparation method for limiting the signal size for feature extraction. In other words, the segmentation goal is to find repeating patterns in the ECG signal known as P, QRS, and T waves [5]. The ECG signal is segmented based on this PQRST complex, which represents key events within each cardiac cycle. These segments, which include the P wave, QRS complex, and T wave, correspond to specific physiological processes in the heart. Fig. 4 (left side) shows the graphical representation of (a) raw ECG signal from the PTB database which consists of noise, followed by (b) filtered signal and (c) shows the 136 different heart beats of P, QRS, T segments. The 290 ECG signals from the PTB database were segmented into 37,752 heartbeats. Similarly, ECG signals from three other databases were also segmented into the respective number of heartbeats. Each individual's ECG signal exhibits a unique heartbeat signature characterized by the PQRST complex making it reliable for biometrics. Fig. 4 (right side) illustrates the variation in heart beat rhythm among four different subjects in terms of shape and peak amplitude, in (a) the amplitude peak is at 0.1, whereas for (d) the peak is at 0.4 and for (c) the peak is at 0.75. These unique properties can be used to derive several characteristics of the heart that helps in extracting features to implement biometrics authentication. Each ECG heartbeat in this study was segmented based on detected R-peaks $r_i$, with a fixed-length window of $L=L_p+L_f+1$ (Lp samples before the peak, Lf samples after it, plus the peak itself). This can be formalized as below equation 1 and illustrated in Fig. 5, followed by equation 2 which shows how peak to peak segmentation can be performed on PPG signal as illustrated in Fig. 6. ECG segmentation leverages the distinct P-QRS-T complex, which provides multiple identifiable fiducial points for analysis. These landmarks enable extraction of a richer set of physiological and temporal parameters compared to PPG, which primarily relies on peak-to-peak segmentation. The multi-feature nature of ECG makes it more robust against noise and variability, enhancing the stability of biometric templates. In contrast, PPG offers fewer distinctive markers, limiting the diversity of features that can be reliably extracted. As a result, ECG segmentation demonstrates stronger potential as a biometric system due to its greater uniqueness and discriminative power.

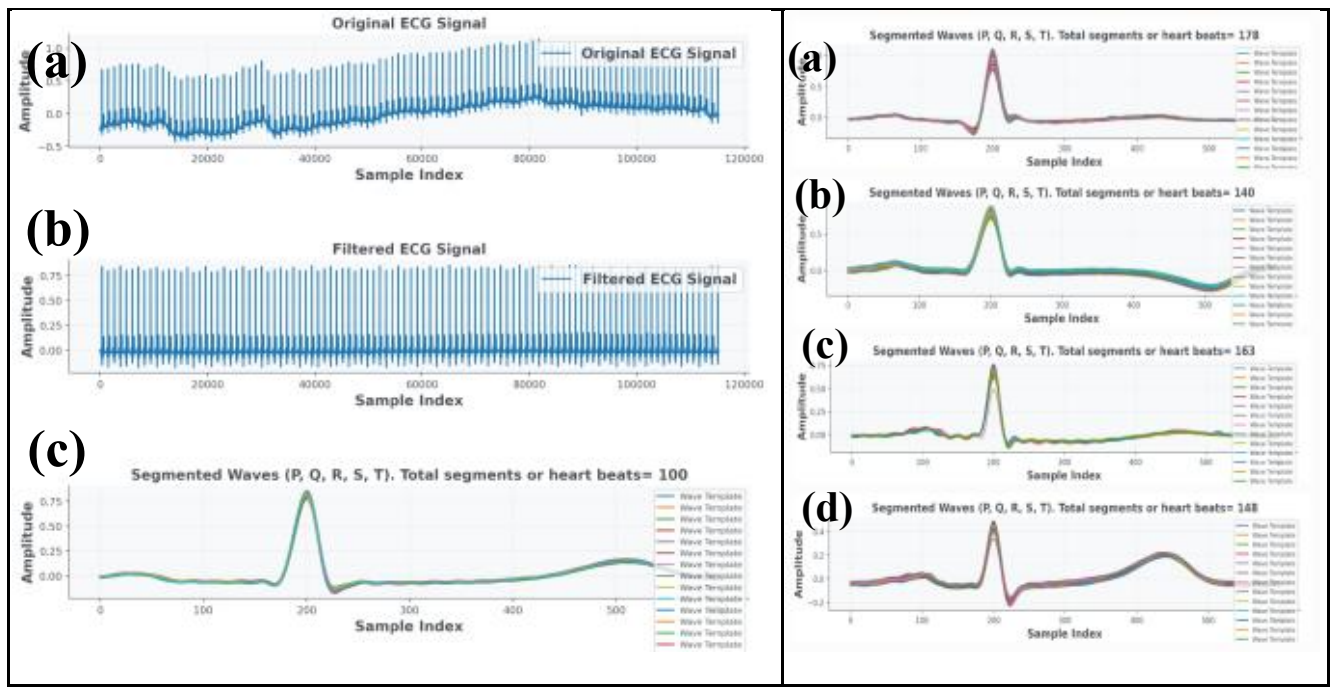

Fig. 4. Graphical representation of Preprocessing stages on the left and illustration of unique rhythm of 4 subjects pulse segmentation.

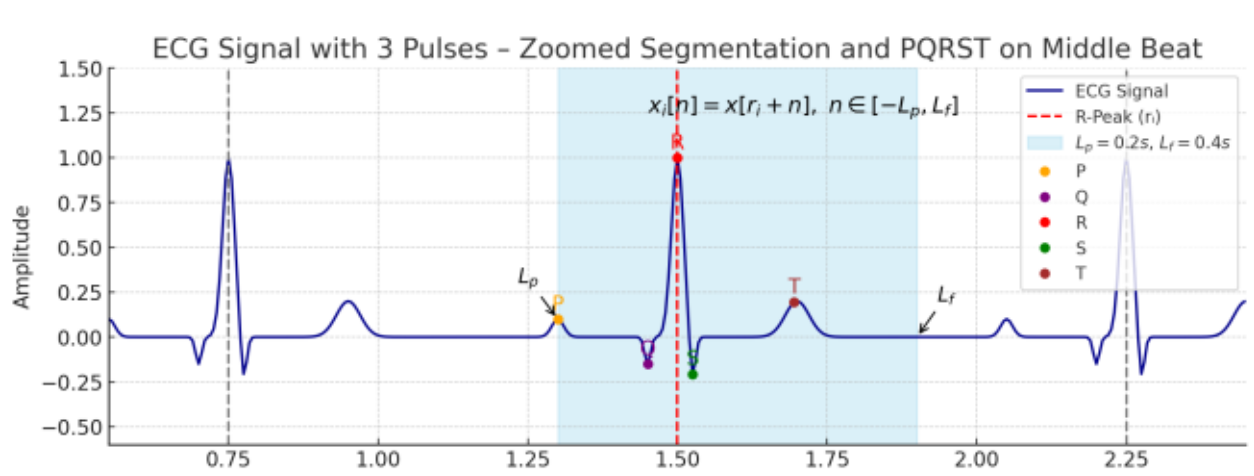

Fig. 5. ECG segmentation equation representation

Eq(1): $x_i[n] = x_i[r_i + n],\ n \in [-L_p, L_f],\ i = 1,2, \dots, N$

Where:
X[n] : Filtered ECG signal
$r_i$: Index of the $i^{th}$ R-peak
$X_i[n]$: $i^{th}$ heartbeat segment
$L_p$: Number of samples before the R-peak (pre-R window)
$L_f$: Number of samples after the R-peak (post-R window)
N: Number of detected heartbeats (i.e, length of R peaks)

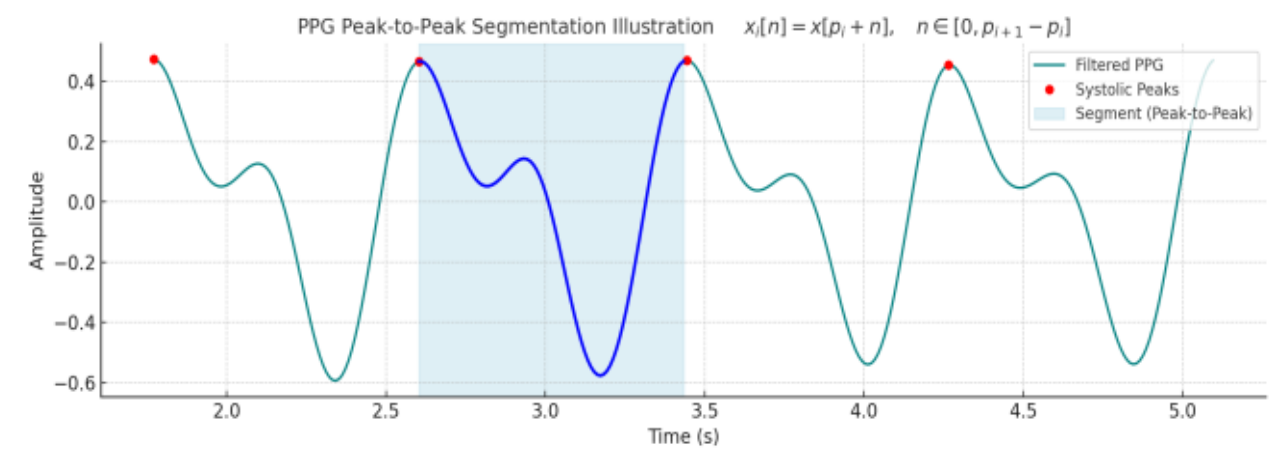

Fig. 6. PPG segmentation equation representation

Eq(2): $x_i[n] = x[p_i + n],\ n \in [0, p_{i+1} - p_i],\ \ i = 1,2, \dots, N$

Where:
X[n]: The original filtered PPG signal as a discrete-time sequence
$X_i[n]$: The $i^{th}$ segmented PPG pulse (one complete cardiac cycle)
$P_i$: Index of the $i^{th}$ systolic peak in the PPG signal
$P_{i+1}$: Index of the $(i+1)^{th}$ systolic peak
n: Time index within the segment
N: Number of detected pulses (i.e, count of systolic peaks)

TABLE II
PRE-PROCESSING COMPARISON BETWEEN ECG AND PPG SIGNALS

| Aspect | ECG Processing | PPG Processing |
|---|---|---|
| Filter Type & Specs | 4th-order Butterworth (Bandpass for 200Hz; Low-pass for higher Fs) | 4th-order Butterworth Bandpass Filter |
| Sampling Frequency | 200Hz, 360Hz, 1000Hz | 100Hz |
| Filter Cutoff Range | [20] 0.5–40 Hz (bandpass for 200Hz); [21] low-pass 40Hz for 360Hz and 1000Hz | 0.5–4 Hz |
| Filter Purpose | Remove baseline wander, body motion, high-frequency noise; preserve PQRST | Remove baseline drift, motion artifacts, high-frequency noise; preserve cardiac waveform |
| Segmentation Strategy | [5] R-peak centered segmentation using PQRST complexes | Peak-to-Peak, Windowing, Peak Detection (Pulse-centered) |
| Segmentation Description | Segments contain P, QRS, T waves representing full cardiac cycles | Segmentation based on systolic peaks, fixed-time windows, individual pulses |
| Variability Handling | Standard segmentation via fixed-length windowing (Lp, Lf) across subjects | Normalized segment lengths to accommodate heart rate and inter-session variability |
| Unique Value | Each individual's PQRST morphology is consistent and identity-distinctive | Individual-specific waveform shape serves as robust biometric trait |
| Segmentation Outcome | Clear amplitude & shape differences among subjects | Peak-to-peak segmentation produced best results across subjects and sessions |
| Databases Used | [15] PTB (1000Hz), [16, 17] MIT-BIH (360Hz), [19] UofTDB (200Hz) | BioSec.Lab PPG dataset (100 Hz) (86 subjects, 2 sessions, 17-day interval) |

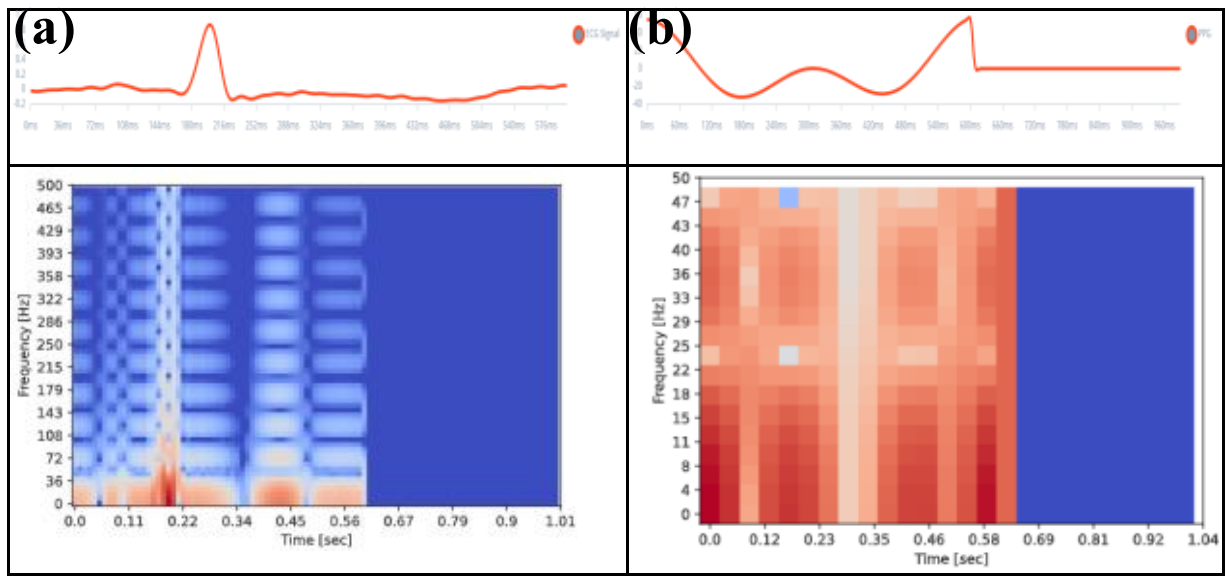

Fig. 7. Spectrogram version of a) ECG, b) PPG segments

## C. Feature Extraction

ECG signals have emerged as a promising biometric for authentication, primarily due to their unique physiological properties. Each individual's ECG is shaped by various intrinsic factors, such as heart size, position, and autonomic nervous system responses, making these signals highly person-specific. The uniqueness of ECG, combined with the difficulty of replicating or forging these signals, positions ECG-based biometrics as a secure and reliable method for authentication. ECG signals exhibit both short-term and long-term variability, offering a rich source of distinctive biometric signatures. Previous studies have utilized various ECG feature categories for biometric identification and authentication. Spectral features derived from the frequency domain of ECG signals have demonstrated significant potential due to their robustness against noise and signal variability [23].

### 1) Segmented Signal Features

From the ECG segmented signals, we can extract detailed time-domain features, such as the timing and amplitude of fiducial points that are critical for capturing the physiological and morphological characteristics of the heart. Additionally, non-linear and complex measures, such as entropy and turning points, can be derived from these segments to provide a deeper understanding of the signal's dynamics. Analyzing these segments allows for a more granular approach to feature extraction, which can enhance the accuracy and reliability of biometric systems. We extracted these features listed in Table III for our analysis and recorded the latency and RAM usage for the extraction along with their importance in determining the accuracy.

### 2) Spectrogram version of segmented signal features

A spectrogram is a visual representation of the frequency content of a signal over time. By converting the segmented ECG signals into spectrograms, we capture both the temporal and frequency information of the signal. This method involves analyzing how the power or amplitude of different frequency components evolves across the PQRST segments. The spectrogram provides a detailed time–frequency representation of the signal (0.02 s frame length, 0.01 s stride, 128-point FFT, −52 dB noise-floor normalization), enabling extraction of features robust to noise and temporal variations. These spectrogram-derived features are valuable for capturing the unique characteristics of an individual's ECG, making them highly effective for biometric authentication [24, 25]. Spectrograms also help in modelling consistent PPG beat-to-beat variations characteristics regardless of small shifts in peak locations [31]. Fig. 7 presents a time vs. frequency plot of spectrograms for ECG and PPG segmented signals, where the color or intensity represents the amplitude or power of each frequency component at each point in time, that are different from one another.

### 3) Features Categories:

To thoroughly evaluate ECG's utility for biometric authentication, we explored 13 distinct categories of features derived from both segmented ECG signals and their corresponding spectrograms. These categories include:

| | | | |
|---|---|---|---|
| I | Time Domain | VIII | Temporal Features |
| II | Frequency Domain | IX | Energy Features |
| III | Statistical Features | X | Peak Features |
| IV | Complexity Measures | XI | Signal Morphology |
| V | Fiducial Points | XII | Harmonic Features |
| VI | Non-linear Features | XIII | Bark Energy Features |
| VII | Spectral Features | | |

### 4) Performance and Reliability of Feature Sets

The comprehensive evaluation of feature categories, as detailed in Table III and IV, highlights the varying computational demands like time taken to extract features from a single segment, memory used to compute the same, and the significance of different ECG features when utilized in an SVM classifier by computing the permutation importance from the scikit-learn library. The tables reveal that certain features, particularly those derived from spectrogram analysis, require more RAM and exhibit higher latency due to their complex computations. Notably, a new feature, turning_points_count the number of times a signal changes direction demonstrated significant contribution toward improving accuracy as shown in Table III. It is crucial to recognize that the combination of features used in the SVM classifier affects overall feature importance. Different combinations can lead to variations in which features are deemed most critical, as turning_points_count showed different performances in Table III and in Fig. 8 on the same dataset a phenomenon documented in earlier studies emphasizing the context-dependent nature of feature importance in machine learning models [26, 27]. This suggests that the selection of feature sets should be tailored to the specific application and available resources. Tables III and IV collectively illustrate that while some feature sets demand more computational resources, they also offer enhanced reliability and accuracy. For TinyML devices, where computational resources are limited, we selected a balanced feature set for efficient computation and performance, as shown in Fig. 8. By carefully combining features, as suggested in prior research [28], we optimize resource usage while maintaining reliable ECG-based authentication for TinyML environments.

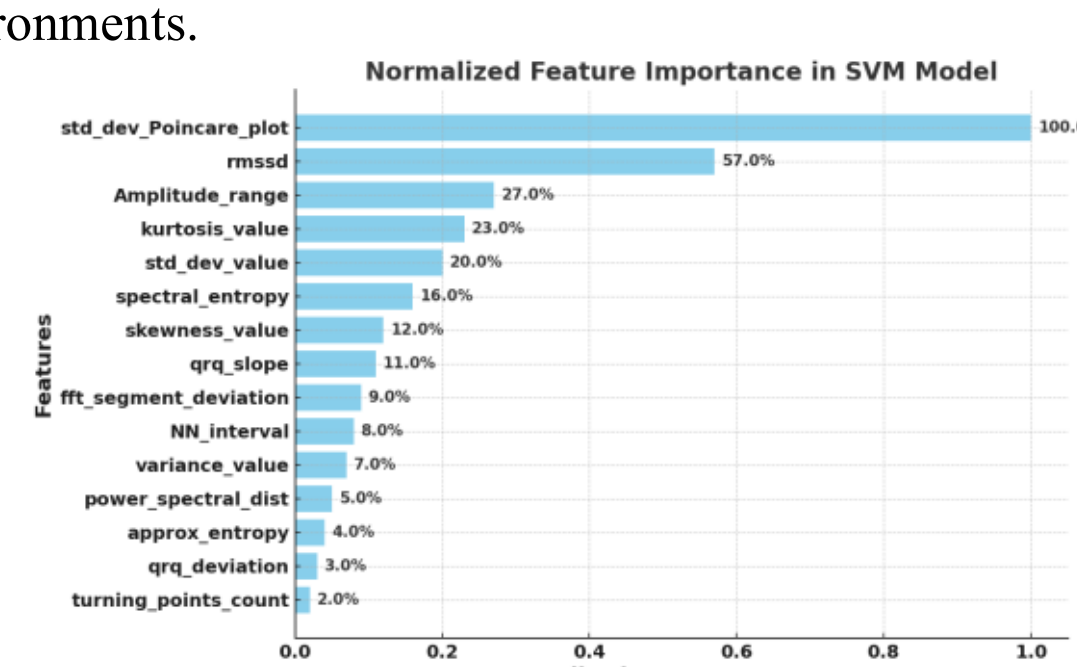

Fig. 8. Combination of important features from Table III that still achieved an overall accuracy of 98%

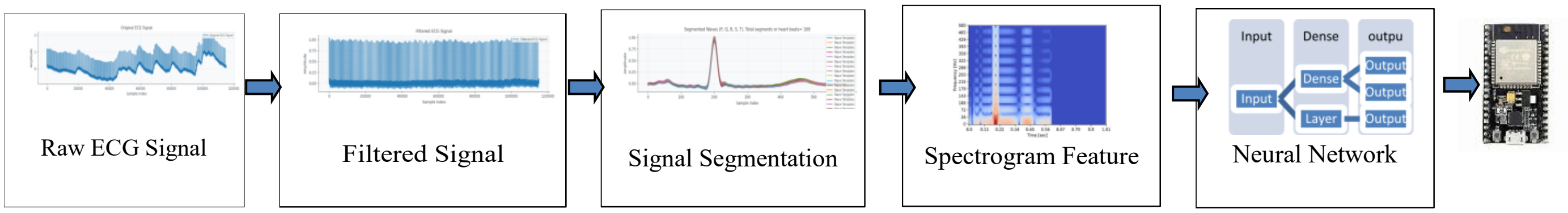

Fig. 9. Flow of proposed methodology implementation

### D. Machine Learning Models for ECG Feature Analysis

We designed the SVM model to understand the behaviour of different features in terms of ECG classification and to lay a benchmark. Later, we developed DNN, 1D CNN, and 2D CNN ML models to compare the TinyML model performances with regular ML models in terms of accuracy and RAM usage.

#### 1) SVM Model Training and Testing

SVMs are popular supervised learning models widely used for classification and regression tasks in various fields. SVMs are particularly well-suited for problems with complex decision boundaries and low-dimensional feature spaces. SVMs aim to find the optimal hyperplane that best separates the feature set points of different ECG signals by maximizing the margin distance between them, and the extracted feature set data points of each class nearest to the hyperplane are known as support vectors. The extracted feature set is labeled to the corresponding ECG signal and provided to a linear kernel SVM model to randomly consider 80% of segmented data of each ECG signal for training and the remaining 20% for testing, where all ECG segments of a subject were assigned exclusively to either the training or the testing set. We evaluated the model using the segmented ECG signal features that are listed in Table III and reported an accuracy of 99%. With segmented spectrogram signal features, which are listed in Table IV, we observed accuracy in the range of 93–94%.

TABLE III
SEGMENTED SIGNAL, FEATURE PERFORMANCE

| Features Extracted | Latency (ms) | Ram (KB) | Feature Importance | Category |
|---|---|---|---|---|
| Turning Points | 0.06 | 2.72 | 62.73% | IX |
| RMSSD | 0.05 | 2.55 | 58.29% | I |
| Std dev in Poincaré plot | 1.36 | 24.4 | 51.05% | VI |
| Kurtosis value | 1.58 | 7.89 | 5.85% | III |
| Multiscale Entropy | 0.06 | 0.98 | 3.81% | IV |
| Baseline relative power | 0.04 | 2.56 | 3.28% | IX |
| P wave Amplitude | 0.02 | 0.97 | 3.03% | V |
| Skewness Value | 1.74 | 7.87 | 1.73% | III |
| ST segment deviation | 0.04 | 0.88 | 1.58% | V |
| Spectral Skewness Kurtosis | 0.31 | 2.06 | 1.47% | VII |
| Median Value | 0.20 | 4.96 | 1.26% | III |
| QRS Area | 0.11 | 3.96 | 0.95% | XI |
| Amplitude Range | 0.04 | 0.86 | 0.92% | XI |
| QRS Slope | 0.11 | 3.96 | 0.84% | XI |
| Sample entropy | 0.33 | 2.83 | 0.71% | IV |
| Spectral Entropy | 0.03 | 1.89 | 0.54% | III |
| Variance value | 0.09 | 2.78 | 0.48% | III |
| STFT Mean | 1.13 | 23.5 | 0.27% | VII |
| Standard Deviation | 0.09 | 2.78 | 0.25% | III |
| Zero Crossing Rate | 0.04 | 2.70 | 0.18% | I |
| P Wave duration | 0.01 | 0.07 | 0.16% | V |
| Mean Value | 0.03 | 0.78 | 0.07% | III |
| QRS Duration | 0.16 | 1.66 | 0.04% | V |
| Power Spectral distribution | 0.81 | 8.74 | 0.01% | II |
| High Frequency | 0.03 | 0.88 | 0.00% | II |

TABLE IV
SEGMENTED SPECTROGRAM FEATURE PERFORMANCE

| Features Extracted | Latency (ms) | Ram (KB) | Feature Importance | Category |
|---|---|---|---|---|
| Spectral Centroid | 0.67 | 1272.9 | 41.96% | VII |
| Spectral Bandwidth | 1.24 | 1273.8 | 41.33% | VII |
| Bark Energy 6 | 12.04 | 2630.6 | 35.46 | XIII |
| Bark Energy 5 | 11.9 | 2636.3 | 35.18% | XIII |
| Bark Energy 4 | 9.47 | 2630.6 | 34.67% | XIII |
| Spectral Entropy | 1.75 | 2267.04 | 26.03% | III |
| Temporal Entropy | 1.79 | 2267.4 | 26.03% | VIII |
| Bark Energy 1 | 11.22 | 2633.3 | 16.94% | XIII |
| Bark Energy 3 | 9.44 | 2630.6 | 15.34% | XIII |
| Max Spectrogram | 0.09 | 1.27 | 14.74% | X |
| Bark Energy 2 | 9.70 | 2630.5 | 13.15% | XIII |
| Std Spectrogram | 0.54 | 1134.9 | 12.19% | III |
| Spectral decrease | 0.95 | 3401.03 | 11.05% | VII |
| Spectral Flatness | 2.53 | 2342.76 | 11.01% | VII |
| Mean Peak height | 2.39 | 1137.8 | 9.55% | X |
| Variance Spectrogram | 0.45 | 1134.9 | 6.24% | III |
| Total Energy | 0.15 | 65.3 | 5.59% | IX |
| Spectral roll off | 2.84 | 1190.3 | 4.59% | VII |
| Mean Spectrogram | 0.16 | 1.17 | 3.59% | III |
| Min Spectrogram | 0.08 | 1.27 | 1.64% | III |
| Spectral Contrast | 0.62 | 86.2 | 0.00% | VII |
| Harmonic Ratio | 0.44 | 75.7 | 0.00% | XII |
| Peak Prominence | 1.89 | 1169.2 | 0.00% | X |
| Temporal Variance | 0.37 | 66.17 | 0.00% | VIII |
| Spectral Flux | 1.02 | 43.49 | 0.00% | VII |

#### 2) NN Model Using Edge Impulse

In Phase-II analysis, DNN, 1D CNN, and 2D CNN models were trained to evaluate their performance with spectral and spectrogram features, designed by Frame Length captures fine details including individual QRS complexes. Frame Stride ensures overlap between frames, maintaining continuity. Suitable for capturing transient events and overall trends. Fast Fourier Transform (FFT) Length corresponds to 64 frequency bins. ECG signals' primary frequency content lies within this range. Normalization removes low-amplitude noise, crucial for handling baseline drift and noise in ECG signals and spectral features extracted by applying the FFT algorithm to each segment of the ECG signal. FFT is an algorithm that efficiently computes the discrete Fourier transform (DFT) of a sequence or signal to fetch a magnitude spectrum, which shows the amplitude or power of each frequency component in the signal. The NN architectures go as below.

I) DNN Architecture

A simple DNN model, which is a linear stack of layers that is created by passing a list of layer instances to the construction, is implemented with two dense (hidden) layers to facilitate model training. Dense layers are the regular deeply connected neural network layers. In this approach, for spectrogram feature analysis, the input layer accommodates 6,435 features representing various frequency parameters of the signal over time for each ECG class. These features are then processed through two dense layers, with distinct combinations of neurons as specified in [29], across varying numbers of epochs. Subsequently, the output layer comprises the total number of classes (subjects) that each database carries. Whereas for spectral feature analysis, the input layer comprises 13 features, followed by 2 dense layers with 60 neurons each and then connected to the output layer. We maintained a learning rate of 0.0005 throughout the training process to avoid overshooting the classification boundaries, thereby ensuring stable and reliable convergence.

TABLE V
NN'S RESOURCE UTILIZATION METRICS

| Parameter | DNN | | 1D CNN | | 2D CNN | |
|---|---|---|---|---|---|---|
| | Int8 | Float32 | Int8 | Float32 | Int8 | Float32 |
| Latency (ms) | 88 | 194 | 71 | 155 | 544 | 4,112 |
| Ram (K) | 7.7 | 26.5 | 15.7 | 52.2 | 65.5 | 254.4 |
| Flash | 223.6K | 841.9K | 146.8K | 484.6K | 1.9M | 7.6M |
| Accuracy (%) | 97.45 | 97.62 | 97.4 | 97.9 | 97.6 | 98.2 |

II) 1D CNN, 2D CNN

The 1D CNN model begins with an input layer that processes features. These features are reshaped into a reduced number of columns to accommodate the structure of the convolutional layers. The first convolutional layer applies 8 filters with a kernel size of 3, allowing the model to detect localized patterns in the ECG data, such as variations in the QRS complex. A dropout layer with a rate of 0.25 follows, which helps prevent overfitting by randomly deactivating a quarter of the neurons during each training iteration. Subsequently, a second convolutional layer with 16 filters, also with a kernel size of 3, is employed to capture more complex patterns within the data. This layer is followed by another dropout layer with the same 0.25 dropout rate, further enhancing the model's robustness against overfitting. The output from the convolutional layers is then flattened into a single vector, which is fed into a fully connected output layer. This final layer classifies the ECG signal into one of the possible categories, corresponding to different cardiac conditions or signal patterns. In parallel, a 2D CNN was implemented to evaluate the impact of spatial relationships within the spectrogram features on classification performance. The architecture mirrors that of the 1D CNN, with the key difference being the use of 2D convolutional layers. These layers can capture two-dimensional patterns across the spectrogram, potentially leading to more accurate classifications. To further guard against overfitting, the dropout rate in the 2D CNN was increased to 0.5, reflecting the increased complexity of the model. Even though CNN provides similar accuracy as DNN, the resource utilization within CNNs is significantly higher than DNN. Table V reveals the resource utilization of the NN models for the PTB database, highlighting high resource utilization for a 2D CNN.

## III. Analysis and Observations

In this section, we will delve into a detailed analysis of data optimization, model performances, identifying the most effective model based on our observations. We will also explore existing challenges and limitations affecting model performance and discuss strategies for improvement. Furthermore, we will outline potential enhancements to the proposed model to address these challenges and optimize its effectiveness

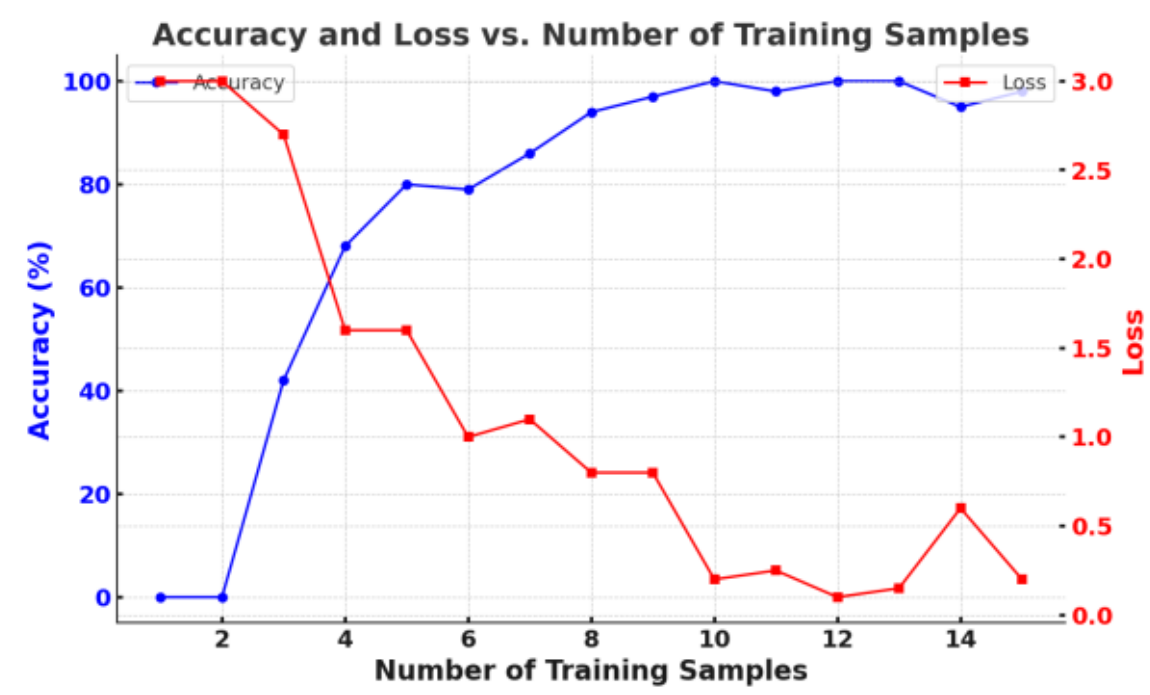

Fig. 10. Model performance for different no. of training pulses

### A. Data Optimization

We addressed DNN model optimization in our previous work [29], that talks about the model response with different DNN architectures. Here, we want to talk more about model performance with a minimum number of training and testing heart pulses. To build a more optimized model, we evaluated its performance using ECG data from 50 different subjects in the PTB database. By testing various combinations of training and testing segments, which is crucial while dealing with a large dataset in real time, ensuring that even with a minimal yet sufficient dataset, the model can generalize well and perform effectively. As visualized in Fig. 10, the performance of our DNN model for ECG biometrics significantly improved with an increase in the number of ECG heart pulses used for training. Initially, with minimal training (1–2 ECG pulses), the model accuracy was extremely low at 0%, with a high loss of 3.06 derived from the categorical cross-entropy loss function. However, as we increased the training samples to 3 ECG pulses, the model's accuracy improved to 41.7%, and the loss decreased to 2.77. This trend of increasing accuracy and decreasing loss continued as more ECG pulses were included in the training set. After training with 4 ECG pulses and testing with 1, 2, and 3 ECG pulses, the accuracy rose to 68.8%, which continued the same trend further as highlighted in green, and the loss dropped to 1.44, highlighting the model is able to demonstrate the same accuracy while testing with just a single ECG pulse. Further training with more ECG pulses resulted in even greater accuracy, with the model achieving 85.7% accuracy at a loss of 0.56 after training with 7 ECG pulses. Our results show that with sufficient training, the model achieved remarkable accuracy. The highest accuracy of 100% was observed when the model is trained with 10 ECG pulses, with a minimal loss of 0.09, demonstrating the model's robust performance. This peak accuracy was maintained by using more ECG pulses for training with a further reduced loss of 0.04. Beyond the 10th cycle, the model consistently maintained high accuracy levels above 96%, reinforcing the effectiveness of our training methodology for ECG pulse recognition tasks. Notably, having a strong signal can potentially reduce the number of ECG pulses required for training, making it more efficient for real-time applications. These findings underscore the importance of extensive training and signal quality in achieving optimal performance in neural network models for biometric recognition, which is crucial for real-time deployment in medical and security applications.

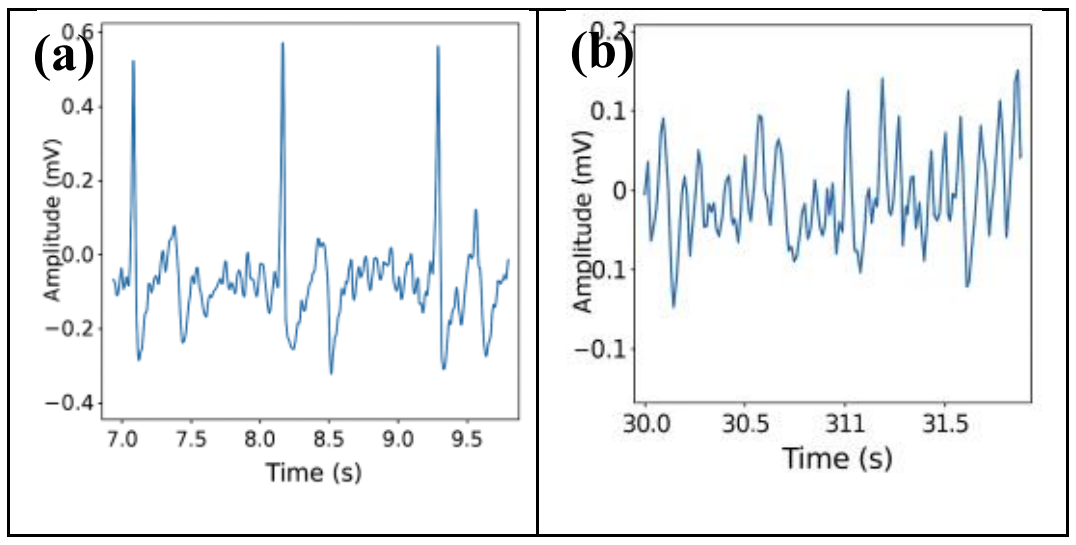

Fig. 11. a) Clean, b) Distorted ECG signal at resting condition

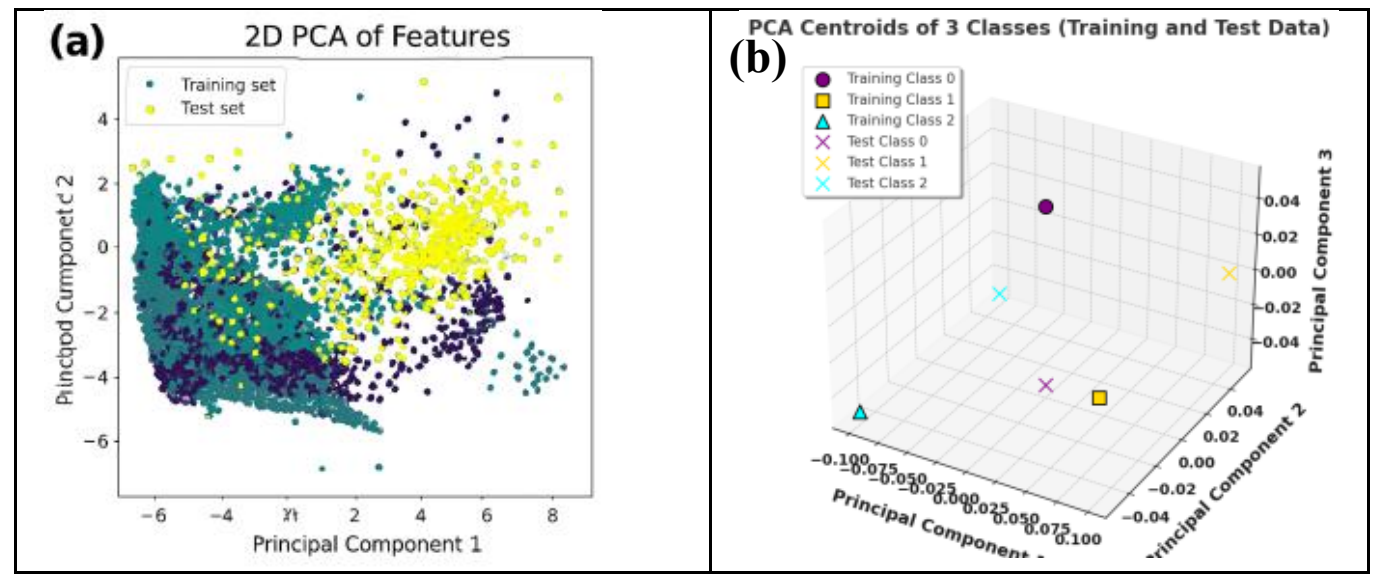

Fig. 12. Visualizing ECG Biometric Identification Space: PCA Projections and Clustering Analysis

### *B. ML Model Observations*

In our analysis of ECG-based biometric identification models, SVMs perform best with lower-dimensional, segmented signal features, offering efficient, robust classification with minimal risk of overfitting by getting trained with features that require more computation, and the performance is reduced for spectrogram version of features (high-dimensional features), as listed in Table VI. In contrast, NN models using one-vs-all classification excel in handling high-dimensional, complex spectrogram features, capturing non-linear relationships crucial for accurate identification. All experiments were repeated five times, and the reported accuracy reflects the average across runs, along with standard deviation to assess model robustness, especially in the presence of signal variability. Performance evaluations across multiple databases highlight these trends, with NNs achieving top metrics on the PTB ECG database, including an AUC-ROC of 1.00 and an F1 score of 0.97. However, challenges like imbalanced data and lower sampling frequency in datasets like UofTDB reveal the importance of high-quality, balanced data for optimal NN performance. Training CNNs for more than 10 epochs increased the computational load, hence we trained CNNs with fewer epochs compared to DNNs, which resulted in the DNN outperforming the CNN model in accuracy. For resource-constrained environments, simpler DNN architectures are more practical than CNNs, balancing accuracy and computational efficiency, especially in TinyML applications.

### *C. Identification Space and Data Analysis*

The concept of the identification space is a multi-dimensional representation where features extracted from ECG data are plotted to visualize the separability and clustering of individual identities. In our previous work [29], we implemented a Uniform Manifold Approximation and Projection (UMAP) based 2D representation of feature distribution across different subjects from the PTB database. For the purpose of visualizing high-dimensional ECG features, Principal Component Analysis (PCA) is employed as a dimensionality reduction technique. PCA transforms the original high-dimensional feature space into a lower-dimensional one while retaining the most critical variance information. This makes it easier to visualize the underlying structure of the data. Fig. 12 illustrates the PCA transformation of spectrogram features extracted from segmented ECG data of the MIT-BIH database. These plots collectively illustrate the identification space of ECG features used for biometric identification via SVM. Fig. 12(a) shows the distribution of features from the training and test sets, highlighting the overall data spread and potential for generalization. Fig. 12(b) visualizes the centroids of three classes, providing insight into the relative positioning and separation of these classes, which is vital for the SVM's ability to form effective decision boundaries.

Also, this shows that while dealing with many subjects, averaging multiple signals enhances error tolerance and accuracy, underscoring the feasibility of ECG-based authentication for secure personal identification in real-time applications along with considering the minimum number of signals required for testing, as shown in Section III under Data Optimization. Together, these visualizations underscore the effectiveness of the extracted features in distinguishing between individuals, which is critical for achieving accurate ECG-based biometric identification.

## IV. Discussion

In this section, we evaluate the overall performance of the proposed model, highlighting its advantages and limitations. A comparison of accuracy results across various machine learning models, as presented in Table VI, indicates that ECG

TABLE VI

Feature Performance with different ML models for ECG data in Different Databases

| Database | Condition | # subjects | Signal Length | Samp freq | Butterworth Filter | Segmented signal features accuracy % with | | | | Spectrogram signal features accuracy % with | | | |
|---|---|---|---|---|---|---|---|---|---|---|---|---|---|
| | | | | | | **SVM** | **DNN** | **1D CNN** | **2D CNN** | **SVM** | **DNN** | **1D CNN** | **2D CNN** |
| **PTB** | Healthy & Cardiac Disorder | 290 | 30 Sec – min | 1000 Hz | Low-Pass (40 Hz Cutoff) | 99 | 34 | 36.6 | 52.5 | 93 | **97.8** | 97.9 | 98.2 |
| **MIT-BIH Arrhythmia** | Arrhythmia | 48 | 30 min | 360 Hz | Low-Pass (40 Hz Cutoff) | 99 | 83.6 | 53.3 | 51.0 | 93 | **95.8** | 95.3 | 94.2 |
| **MIT-BIH Stress** | Stress Test | 28 | 30 min | 360 Hz | Low-Pass (40 Hz Cutoff) | 99 | 88 | 71 | 68.5 | 94 | **97.3** | 98 | 98.0 |
| **UofTDB** | Sit | 1,012 | 2-5 min | 200 Hz | Bandpass (0.5-40 Hz) | 92 | 65 | 53.5 | 55 | 75 | **91** | 90 | 91 |
| | Stand | 81 | | | | 94 | 70 | 58.5 | 58.7 | 80 | **94.9** | 91.5 | 91.5 |
| | Exercise | 71 | | | | 87 | 66 | 35.8 | 38 | 70 | **85** | 78.6 | 80.2 |
| | Supine | 63 | | | | 95 | 73 | 60 | 62.3 | 90 | **95.8** | 95 | 96 |
| | Tripod | 63 | | | | 96 | 72 | 62 | 66 | 93 | **97** | 95 | 96 |

data with high-frequency components, such as those from the PTB database, contain richer information, making them highly suitable for classification, even when the ECG data encompasses diverse health conditions. This is particularly true when employing a spectrogram-based classifier. Moreover, the model demonstrates consistent performance when applied to arrhythmia data from the MIT-BIH Arrhythmia Database, reinforcing its reliability under pathological conditions. The model's robustness is further validated in the MIT-BIH Stress Database, where it maintains stability even in the presence of real-world noise. However, a slight decline in performance is observed with the UofTDB dataset, underscoring the importance of training with high-quality, high-frequency signals. This strengthens previous studies that have shown that while a sampling rate of 120 Hz may be sufficient for general heart rate monitoring, a higher frequency of around 360 Hz is preferable for more advanced ECG analysis [30]. Fig. 11 illustrates the difference between clean and distorted ECG signals under the same conditions, as collected from UofTDB. The distortion makes it challenging to identify the P, QRS, and T parameters required for symmetrical segmentation, complicating the modelling process for detecting similar feature patterns.

Despite this, the proposed model effectively handles slight variations in ECG signals due to different physical activities, enhancing its reliability across various scenarios after training. Among the machine learning models and feature sets explored, the DNN model utilizing spectrogram-based segment features consistently outperforms others across different ECG databases. This combination proves to be the most effective for classification tasks. Additionally, adapting the model for TinyML devices broadens its applicability, enabling integration into advanced applications such as wearable devices and personal electronics, including mobile phones, laptops, and beyond. In contrast, similar research [32] found that PPG signal performance declined as the dataset size increased, underscoring the relative advantage of ECG-based methods for scalable biometric authentication.

## V. Conclusion

In summary, this journal explores the substantial potential of a robust TinyML-based DNN model for utilizing ECG signals as biometric identifiers on resource-constrained devices. This study delves into the development of a TinyML model for ECG biometrics, assessing its performance across diverse ECG datasets under varying health conditions, noise levels, sampling frequencies, and situational contexts, achieving exceptional accuracy rates ranging from 91% to 98% across different sessions accuracy that could be improved with higher sampling frequencies and distortion-free ECG data. Additionally, we evaluated the efficacy of different feature sets under spectrogram and spectral features, utilizing SVM, 1D CNN, 2D CNN machine learning models in conjunction with these features. Also in this work, we introduced the concept of the identification space, which optimizes biometric authentication for real-time applications by requiring only 10 ECG segments for training and a single ECG segment for authentication. This is important in real-time applications where processing time is critical. Moreover, ECG pulses contain a richer set of parameters, providing more distinctive biometric information, which makes ECG inherently more robust for large-scale authentication compared to PPG-based spectrograms. This approach enhances efficiency, making the model highly practical for implementation on resource-constrained devices, making it highly suitable for real-time, low-latency applications such as wearables and mobile devices. Integrating biometrics with TinyML for remote biometric functionality presents a powerful combination of security, efficiency, privacy, and scalability. The application of TinyML-based ECG signal biometric analysis paves the way for numerous innovations, particularly in fields such as military technology, wearable devices, and healthcare. Future research can be extended by studying feature behavior to distinguish between synthetic and natural ECG, testing the model with real-time ECG data collected under various states of physical and mental stress, anxiety, and by designing a wearable device capable of measuring and sensing ECG signals through a sensor, with real-time classification performed by TinyML, that can be implemented into daily wearable devices like smart rings, watches, mobile devices, jackets, and wrist bands.

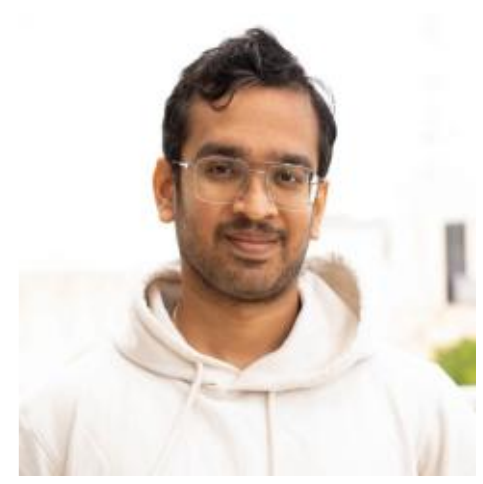

**Yogeswar Reddy Thota** (SM'24) received the bachelor's degree in Electronics and Communication Engineering from Annamacharya University, Andhra Pradesh, India in 2021, and pursuing master's degree in computer engineering in the University of Texas at Dallas, TX USA (2024-25). He is currently focused on developing Machine Learning models for edge devices, with research interests spanning TinyML, Edge AI, Large Language Models (LLM), Small Language Models (SLM), and Data Engineering.

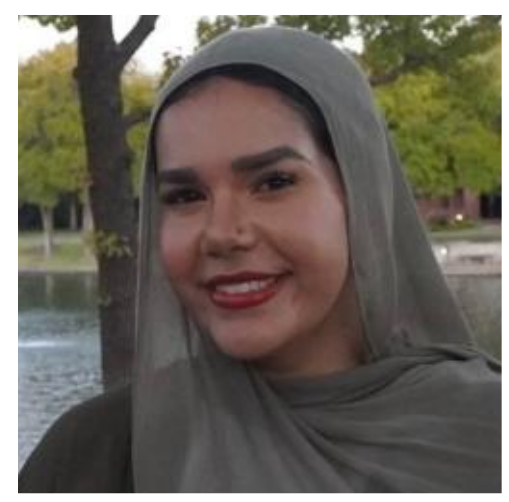

**Hanan Alsmadi** (SM'24) is pursuing a Bachelor of Science in Electrical Engineering from the University of Texas at Dallas, Dallas, TX, USA. She is currently focused on studying the applications of TinyML and enabling these devices to process data locally. Her academic interests include embedded systems, signal processing.

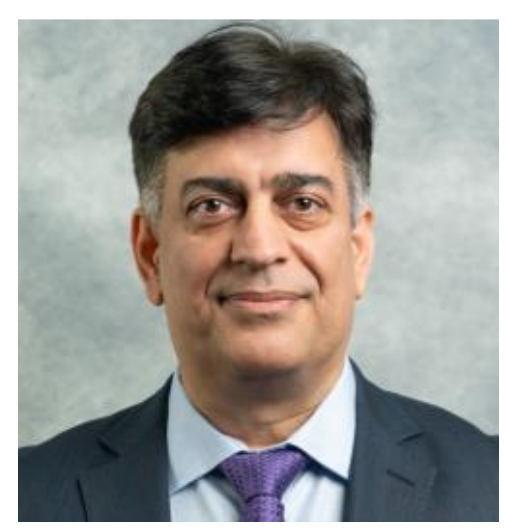

**Tooraj Nikoubin** (SM'15) received his B.Sc. and M.Sc. degrees in Electronic Engineering from the Department of Electrical and Computer Engineering at K.N.Toosi University of Technology, Tehran, Iran, in 1991 and 1994, respectively. He earned his Ph.D. in Computer Engineering from Shahid Beheshti University, Tehran, in 2009. From 2012 to 2019, Dr. Nikoubin was a faculty member in the Department of Electrical and Computer Engineering at Texas Tech University. In 2019, he joined the University of Texas at Dallas as a Professor in the Department of Electrical and Computer Engineering. His current research interests include Embedded AI and Smart Sensors, Low-power and Energy-efficient Design and Optimization, Hardware and Machine Learning for edge devices, Approximate Computing, and Wearable Electronics for health and assistive technologies.